\documentstyle[12pt,aaspp4]{article}
\def\farcs{\hbox{$\> .\!\!^{\prime\prime}$}}

\begin{document}

\title{{\it HST} Imaging of the BL~Lacertae Object OJ~287\altaffilmark{1}
\altaffiltext{1}{Based on observations with the
NASA/ESA {\it Hubble Space Telescope}, obtained at the Space Telescope 
Science Institute, which is operated by the Association of
Universities for Research in Astronomy, Inc., under NASA contract
NAS5-26555.}}

\author{Brian Yanny}
\affil{Fermi National Accelerator Lab., Batavia, IL 60510, USA}
\affil{Email: \tt yanny@sdss.fnal.gov}

\author{Buell T. Jannuzi}
\affil{National Optical Astronomy Observatories, P.O. Box 26732,
Tucson, AZ 85726-6732, USA}
\affil{Email: \tt jannuzi@noao.edu}

\author{Chris Impey}
\affil{Steward Observatory, University of Arizona, Tucson, AZ 85721, USA}
\affil{Email: \tt impey@as.arizona.edu}

\begin{abstract}

{\it Hubble Space Telescope} WFPC-2 $I$-band (F814W) images of the
BL~Lacertae object OJ~287 and the surrounding field are presented.  We
find evidence of associated extended nebulosity near OJ~287, as well 
as a small nebulosity to the West, which may be spatially 
coincident with the position of previously observed radio emission.  
The brightness of a host galaxy is difficult to determine due to the 
brightness of the active nucleus, but it lies in the range $-21.5
> M_R > -23.1$ ($H_0$ = 100 km s$^{-1}$ Mpc$^{-1}$, $q_0 = 0$). No 
evidence is seen for the previously reported optical ``jet'' at 
position angle 220$^\circ$ to a surface brightness limit of $I = 
24.3$ mag arcsec$^{-2}$.  There are several resolved and unresolved
objects within 17$''$ of OJ~287 in the field to limits of $I=25$ (point
source 5$\sigma$ detections). The magnitudes and relative positions 
of these objects are reported.  An offset in the centroid position 
between the OJ~287 point source and the underlying nebulosity reported 
by Wurtz, Stocke and Yee is confirmed and 
measured to be about 0\farcs 4 (1.2$h^{-1}$ kpc at the redshift of OJ~287).  
This offset is tentatively interpreted as evidence for recent merger 
activity rather than a sign of gravitational microlensing. 

\end{abstract}


\section{Introduction}
 
BL~Lacertae objects are compact radio sources that show clear manifestations
of nonthermal activity: power law energy distributions, and rapid flux and
polarization variability.  Rapid flux variability, one of the defining 
characteristics of BL~Lac objects, is due to the changes in observed
luminosity of the synchrotron emission produced from the bright
unresolved source that dominates the observed emission from the core 
(Wagner \& Witzel 1995).  Believed to be generated in a ``jet'' of 
relativistic material viewed nearly along the jet axis (Blandford \& 
Rees 1978), the synchrotron radiation greatly complicates efforts to 
observe the nearby environment of BL~Lacs.

Much current research on BL~Lac objects focuses on the properties of
the active nucleus across the electromagnetic spectrum. There has also 
been interest in the host galaxies and environments of BL~Lacs 
(e.g. Falomo 1996; Wurtz et al. 1997).  A goal of direct observations of BL~Lac host 
galaxies and companions is a better understanding of the conditions 
necessary for the generation of the BL~Lac phenomenon, specifically the 
role that galaxy-galaxy interactions might play in triggering and 
maintaining the activity in BL~Lacs. 

The largest ground based optical survey of BL~Lac hosts to date is 
that of Wurtz, Stocke, \& Yee (1996, hereafter WSY), with 50 objects 
imaged.  A host is resolved in over 90\% of the objects, and based on
radial profiles, at least 70\% (possibly as high as 90\%) show 
elliptical type host galaxies, with no more than 12\% showing 
exponential disk or spiral type hosts. Studies using the higher 
resolution available to the Hubble Space Telescope ({\it HST}) have also 
begun, with advantages for seeing the host galaxy close to the nucleus,
and for detecting close companions. Falomo et al. (1997) have presented
results on three objects; two have elliptical hosts and one does not 
show a resolved underlying host to faint limits.

The optical variability of the well known BL~Lacertae object OJ~287, at 
$z = 0.306$, has been monitored from the ground more extensively than 
any other BL~Lac object (see, for example, Sillanp\"a\"a et~al.
1996). Also known as $0851+202$, OJ~287 has extremely variable total and
polarized flux, reaching very high degrees of observed 
optical polarization (Angel \& Stockman 1980). The $B$-band 
brightness varies between $B\sim 12.5$ and $B\sim 18.5$ over time 
scales of years with smaller fluctuations on smaller timescales
(Gonzales-Perez, Kidger, \& de Diego 1996). The observed colors 
of $V-R = 0.5\pm 0.1$ and $V-I \sim 1.4$ are relatively constant despite the
observed variability (Efimov \& Shakovsky 1996).  

Recently Ben\'itez et al.~(1996, hereafter B96) obtained deep 
ground-based optical and IR images of OJ~287 and reported extended 
nebulosity, possible spatial coincidence between extended optical and
radio emission, and a candidate optical ``jet'' (of substantial angular 
and physical size) at a position angle of 220$^\circ$. This putative
optical jet does not coincide with the position angle of the VLBI
radio jet ($\sim 265^\circ$), which is inclined at 
only 15$^\circ$ to the line of sight and is responsible for the 
apparent superluminal motion of the compact
radio source (Gabuzda \& Cawthorne 1996;  Vicente, Chalot, \& Sol 1996). 
The existence of an additional optical jet would be surprising because 
of the expectation that in OJ~287 we have a highly foreshortened 
view of the ejection axis; dual radio and optical jets would also 
be unprecedented in any AGN.

OJ~287 is notable among the relatively modest numbers of BL~Lac objects:
it is among the brightest, the most luminous, the most compact (at radio
wavelengths), and the most rapidly variable in the sky. It is a clear 
candidate for high resolution space-based studies since its bright
point source flux makes an underlying host galaxy difficult 
to resolve from the ground based observations as light imaged from the 
wings of the point source scatters into the host image.
In this paper we present the first {\it Hubble Space Telescope} ({\it HST}) 
WFPC-2 images of the OJ~287 field and discuss what can be learned about 
the host galaxy and environment of OJ~287.

\section{Observations and Data Processing}

The {\it HST} images include two 500 s, two 400 s, and two 30 s WFPC-2
exposures of the target obtained 24 March 1996.  Only results from the
central PC CCD ($r<17''$ from the BL~Lac nucleus, corresponding to
51 $\rm h^{-1}_{100}$ kpc at z=0.306) are presented here.

The four long exposures were cosmic ray cleaned in pairs using the
technique described in Yanny et al.~(1994) and averaged for an
effective total exposure time of 1800 s.  At the time of the
exposures, OJ~287 was near the bright end of its previously observed
range of $I$-band magnitudes, ($I\sim 13.3$), and all long exposures
were saturated by a factor of 20 in the core out to a radius of 4
pixels (0\farcs 17 at the scale of the PC CCD). A combined image is 
presented in Figure~1. Notation for the objects detected near OJ~287
follows and extends the nomenclature of B96. 
The faint horizontal band running across the center of the image is an 
artifact due to heavy saturation in the center of OJ~287 affecting the 
gain of the WFPC-2 PC CCD readout 
electronics (Biretta, Ritchies, \& Rudloff 1995).

Availability of an accurate point spread function (PSF) model is
essential to good subtraction of the OJ~287 point source, and it allows 
statements to be made about residuals. The inner PSF of WFPC-2 is relatively 
stable with time.  The optics of the {\it HST} are well known and a PSF can be
constructed theoretically.  Software to do this exists and is called
``Tiny Tim'' (Krist 1996).  This PSF is relatively accurate out to radii
of $r \sim 1$\farcs 5.  Further out, variable scattered light within the
{\it HST} is impossible to model theoretically and can only be modeled 
empirically.

In Figure~2a we show a Tiny Tim model WFPC-2 PC PSF.  Archival {\it HST}
images of stars may also be used as PSFs and are often found to be
preferable when dealing with a very bright or saturated point 
source.  The difficulty lies in obtaining sufficient signal-to-noise
in the wings while not saturating the core. A single
observed PSF cannot be used to simultaneously fit the intensity of
the core and model the faint wings of the image.  A PSF constructed from
a combination of long and short exposures provides a better model.  Since the PSF varies 
substantially with filter, with intrinsic color of the point source, 
and with position on the CCD over scales of 100 
pixels or less, it has thus far been impractical to build up a good 
library of PSF observations for the entire WFPC-2 camera field in all 
filter combinations.

An extensive search of the {\it HST} archive yielded the  WFPC-2 PC 
archival PSF shown in Figure~2b.  This PSF is located somewhat off-center of 
the PC chip, and has been cosmic ray cleaned.  It has excellent signal-to-noise 
in the wings. It is saturated to $r = 10$ pixels in 
the core and can only be used to subtract wing light from the OJ~287 
point source.

We have subtracted both the theoretical and the empirical
(archival) PSFs from OJ~287. The results were compared in detail, and
while the two residual images agree qualitatively, the
theoretical PSF is a better match.  The quantitative estimates
which follow were derived from a subtraction of the PSF shown in Figure~2a.  
The residual image after subtraction of the theoretical PSF is presented
in Figure~3, with the same orientation as Figure~1.  The exact amount by
which a model PSF must be scaled to match the light in OJ~287 is the key 
quantity.  We were able to use light from the unsaturated pixels
beyond the core pixels to determine a relative scaling for the PSF.  
Additionally, non-azimuthally symmetric extended emission allows us to 
judge an appropriate scaling by searching for a scaling of the model PSF 
which results in a smooth, but not over-subtracted, residual.  The centering 
of the PSF is accurate to better than 0\farcs 04, as experiments which shift 
the model PSF by that amount and subtract result in a noticeably poorer fit.

Figure~3 shows an apparent excess of light extending over several 
arcseconds to the NW of OJ~287 and SW of object E, which we believe
is related to OJ~287, possibly an OJ~287 host galaxy.  This light is not 
centered on OJ~287.  It is difficult to accurately estimate the 
difference in centroids between the point source.  However, based on
the fact that there is a great deal more extended light to the NW
of OJ~287 than to the SE, we estimate that the difference in centroids 
is 0\farcs 4 or 1.2$h_{100}^{-1}$ kpc at $z = 0.306$, the redshift 
of OJ~287 calculated from emission lines (Sitko \& Junkkarinen 1985).  

Several checks have been performed attempting to confirm that this off-center light 
is not an artifact. Both empirical and artificial model PSF light was 
subtracted from strongly saturated stars obtained from the {\it HST} image 
library in order to determine if residual light due to the optics of 
{\it HST} could account for off-center diffuse light such as that seen in 
Figure~3. While some highly saturated stars did show non-azimuthally 
symmetric scattered light due to the {\it HST} optics, none of the scattered light
was in the specific orientation or of an amplitude which approached the level of the
nebulosity seen in Figure~3.  Small centering shifts of 0.5 to 2 pixels 
in the position of the subtracted PSF have no effect on the presence 
or flux of the residual light seen on scales of 20 to 50 pixels 
NW of the BL~Lac.  Also, subtraction of the same scaled PSF from a 
bright field star in a similar {\it HST} image shows no extended 
residual (Jannuzi et al. 1997).

The available evidence indicates that the observed NW extension of the 
faint radiation around OJ~287
is not produced by any telescope or instrument problem, and in fact 
confirms an earlier suggestion by WSY of decentered
nebulosity. Figure~4 of WSY suggests a 0\farcs 1 separation between
the point source centroid position and point$+$extended source centroid. Given 
their ground based resolution, this is consistent with our 0\farcs 4 
measurement of separation between point source centroid and
extended source (alone) centroid. Additionally, the orientation of 
the extended light, significantly to the W of OJ~287, shows that it 
is not light scattered from object E, 3$''$ N of OJ~287 (This point was 
also noted by WSY).  There is a small knot-like 
object (OJ$-$SW) 2\farcs 5 SW of OJ~287, indicated by arrows in 
Figure~3.  This feature is also not an artifact of the PSF subtraction.  

There is known radio emission coincident with object OJ$-$SW, but the radio
emission extends for 8-10$''$ to the W of OJ~287, much further out than
we detect optical emission. We find no clear evidence for an extremely 
faint optical source located $\sim 8''$ W of OJ~287 at $I>$25 in Figure~3.  
This source would correspond to the source noted in the deconvolved 
image of B96 and corresponds in position to the radio emission seen 
in Kollgaard et al. (1992) and Perlman \& Stocke (1994).  The position angle
of the radio ``jet'' to the West, at P.A. 265$^\circ$, agrees with
the P.A. of the parsec scale jet resolved in VLBI observations (Gabuzda and
Cawthorne 1996).  We note that the ground based image of B96 appears 
to reach detection limits about one magnitude fainter than the 
present {\it HST} image, and the object was at the limit of detection in their 
combined raw image.  Deeper optical imaging followup is clearly needed.

The {\it HST} image shown here does not show evidence for
an extended ``jet'' at a position angle of 220$^\circ$ as suggested in B96.  
While from the ground the objects A, B, and C, (and  I and J) appear 
nearly co-linear with OJ~287, in fact, they are clearly not co-linear 
when viewed at the resolution of {\it HST}. Surface brightness limits on binned 
continuum $I$-band light between A, B, C, I, and J are 
placed at SB($I$) $\rm > 24.3 \> mag\> arcsec^{-2}$, nearly as faint as the levels
reached at lower resolution from the ground in B96. 
Our surface brightness limits on the lack of connecting emission 
between OJ~287 and objects at B and C are inferred from a $10\times 10$ pixel 
binned version of the {\it HST} image, in which no hint of an optical jet is 
seen.  

There is a second distinct object located 1$''$ SW of object E 
identified as object OJ-Gh.  The position of this ringed feature on the 
WFPC-2 CCD image and its similarity to known ghost images of bright 
sources in the WFPC-2 {\it HST} optics make it
suspect (Biretta, Ritchies, \& Rudloff 1995).  We believe this object
to be a ghost image of OJ~287 due to reflections in the {\it HST} WFPC-2 optics, 
and do not consider it further.  

In Table~1 we present positions, Kron-Cousins $I$-band magnitudes, an extent
indicator, and ellipticities for objects in the OJ~287 field. 
The $I$-band magnitudes listed in Table~1 are sky-subtracted aperture 
magnitudes calculated according to the formula given in the WFPC-2 
photometry cookbook (Whitmore 1996).  The radius of the aperture is 
10 pixels for small objects, and 150 pixels for OJ~287 itself,
where 1 pixel = 0\farcs 044.  Since only one filter is available,
color corrections between F814W and the $I$-band may introduce errors 
of up to 0.1 magnitudes in the photometry.  The difference
$I_{\rm KC}-F814W_{inst}$ is set to $-$1.2 mags, a typical
value for objects of reddish color as defined in Whitmore (1996).

We calculate upper and lower limits on the brightness of the extended 
emission underlying OJ~287 and to the NW as follows: For a sample 
of several PSF subtractions (both model PSFs and taken from
the {\it HST} archive) of badly saturated objects (images of bright 
isolated stars again taken from the {\it HST} archive), there can be 
residual light left of order about 10\% of the total light 
of the original object, while the wings appear adequately subtracted.  
This 10\% maximal residual light can also be seen in 
the OJ~287 image after minimal PSF subtraction by either the model or empirical 
PSF.  This places an upper limit on the brightness of an OJ~287 host 
galaxy at  $I >$ 15.8.  We can place a stronger upper limit on the 
brightness by noting the azimuthally symmetric properties of the residual light.
Estimating this contribution by studying the residual light 
of saturated stars, the upper bound for the brightness 
of the host galaxy is  $I>$ 17.0.  To determine a lower limit, we note the 
residual light to the NW of OJ~287.  This light has a surface 
brightness of about 0.25 DN pixel$^{-1}$ (SB($I$) = 21.3 mag arcsec$^{-2}$)
above background compared with a patch of sky SE of OJ~287 and diametrically 
opposite of the excess seen to the NW.  This excess, if spread 
uniformly over the area of sky that encompasses the entire area underneath
the obliterated OJ~287 residual, yields a flux corresponding 
to $I<$ 18.6.  For reasonable assumptions about color, this surface brightness is
well within the range of host galaxy surface brightnesses listed in WSY.
If there is no light directly underneath OJ~287, the measured diffuse light 
has an integrated magnitude $I\sim 19$. 

For $q_0 = 0, H_0 = 100 h_{100}$ km s$^{-1}$ Mpc$^{-1}$, these magnitudes 
correspond to host galaxies in the absolute magnitude range from $ -21.5 
> M_R > -23.1$, where $R$-band is the rest wavelength spectral
region probed by $I$-band light at z=0.306.
Our favored estimate for the extended light surrounding 
OJ~287 is $I = 18.3$, and $M_R = -21.8$.

For comparison, WSY estimate from the ground, when the variable point 
source BL~Lac was one magnitude fainter, that the magnitude of the OJ~287 
host is $r = 19.8$ in the Gunn system.  

\section{Discussion}

There are a total of 6 extended source objects 
within $r<10$\farcs 2 ($\sim$30 h$^{-1}_{100}$ kpc) of the BL~Lac. 
It is possible that object E is closely associated with the BL~Lac, and 
several of the extended objects such as F and G have extents and magnitudes
consistent with galaxies at the same redshift as OJ~287.  However, we note
that based on the Hubble Deep Field galaxy counts (Williams et al. 1996),
the number of galaxies expected with $I_{814} < 23.5$ in a circular patch 
of radius 22$''$ is $\sim 7$, consistent with the number of neighbors
 observed here within 10$''$, allowing for a clustering correlation
enhancement.  Although this shows there
is not compelling statistical evidence for believing that any of these
galaxies are physical companions, the objects (OJ~287, OJ-Ext, OJ-SW, A, E, F
and G) are suggestive of an interacting group of objects spread over
an extent 50 h$^{-1}_{100}$ kpc.  The unresolved objects B and C are 
likely to be Galactic field stars of type K or M, based on their multicolor 
photometry from B96.  Wurtz et al. (1997) find no evidence for a rich cluster
surrounding OJ~287.

It is interesting to note the positional coincidences in radio emission 
contours seen in the 6 cm map of Kollgaard et al. (1992) and resolved 
sources A, F and G in Figure~3.  It is not known if these sources 
are physically associated with OJ~287 at z=0.306.   Obtaining redshifts for
these objects will decide the issue, though the redshift of the diffuse
faint object G with I=23.3, will be difficult to obtain.

There is a clear decentering (by at least 0\farcs 4 to the NW) of the point source 
from the main body of extended emission surrounding the BL~Lac.
The observed decentering is consistent with what
might be expected for on-going interaction or merger activity between an 
object containing OJ~287's gaseous fuel supply and the decentered 
galaxy NW of OJ~287. 

Hernquist (1989) demonstrates how a merger or close interaction can cause a 
rapid clumping of the gas supply of a galaxy to serve as fuel 
for an active nucleus.  Interaction-induced activity is particularly attractive 
in the case of OJ~287, because a binary black hole model can account not 
only for the optical flux periodicity (Lehto \& Valtonen 1996), but also 
the VLBI radio morphology (Vicente, Charlot, \& Sol 1996). A merger or 
interaction scenario feeding gas to an active core appears more likely than a 
gravitational microlensing explanation for the high variability and 
decentered host galaxy of OJ~287.   Microlensing scenarios have difficulty
achieving the observed ``variability duty cycle'':  The low optical depth 
of foreground lenses (typically $\tau \sim 10^{-6}$), leads to duty cycles 
where a lensed BL~Lac would go through rapid 30\% brightness variability 
episodes due to microlensing only once or twice per year as clusters 
of stars in a foreground galaxy pass in front of the BL~Lac core
along the line of sight (Treves et al. 1997).  OJ~287 varies much more 
frequently with much higher amplitude brightness variations 
than predicted by straightforward microlensing scenarios. 

If interactions are common and direct a gas supply to feed a BL~Lac central
engine (although we note that an external gas reservoir by no means guarantees
nuclear activity), then more decenterings or additional nuclei near BL~Lacs are 
predicted.  The fraction of objects with observable decenterings may only be a 
few percent, however, if the interacting objects rapidly merge on 
a dynamical timescale. We see no `second nucleus' near 
OJ~287, though several kpc around the point source are unobservable due 
to the brightness of the BL~Lac.  Falomo et al. (1997) and Jannuzi et al. (1997) 
each investigated three BL~Lac host galaxies with the high resolution 
of {\it HST}, and rule out decentering in five objects to levels of 0\farcs 05.  
No extended emission was detected for the sixth object (Falomo et al. 1997).

There are only a few cases of BL~Lacs with confirmed optical jets at
large angles to the line of sight.  The best known case is that of
PKS0521$-$36 (Macchetto et al. 1977).  In the present case of OJ~287,
objects B and C in Figure~1 are not co-linear with objects A and OJ~287 
and are in fact likely to be Galactic stars rather than knots in a 100 
kpc long optical jet (B96). The evidence that OJ~287 belongs to the class 
of rare BL~Lacs with optical jets is still weak.  Deeper images are necessary 
to confirm the claim of B96  of an optical counterpart to the radio 
emission 8$''$ West of OJ~287.

Our best estimate for the absolute magnitude of the off-center host galaxy 
is $M_R = -21.8$, while it could be as bright as $M_R = -23.1$. The radio 
flux at 20 cm from the extended source is log $P$ = 23.3 W Hz$^{-1}$ 
(Perlman \& Stocke 1994).  OJ~287's observed optical and radio emission
places it in the FR~I region of a radio-optical flux diagram, with properties similar to 
many other known BL~Lac hosts.

\section{Conclusions}

Significant progress has been made in the study of the host galaxies
and environments of BL~Lac objects with ground-based observations
(WSY; Falomo 1996). However, these studies have necessarily focused on 
the nearest objects and have not been able to resolve their inner 
regions.  {\it HST} observations presented here show 
the BL~Lac OJ~287 to be interesting in several respects:

\begin{itemize}

\item [1.]{Either the OJ~287 host galaxy is decentered by at least $1.2 h^{-1}$ 
kpc from OJ~287 itself, or there is an interacting companion galaxy to 
an unseen OJ~287 host located at small projected separation from OJ~287.}

\item [2.]{The off-center nebulosity has an elliptical shape, absolute 
magnitude in the $I$ (redshifted $R$) band of $-21.8 > M_R > -23.1$,  and a 
surface brightness at one effective radius of 
SB($I$) = 21.3 mag arcsec$^{-2}$, consistent with many known FR~I type galaxies.}

\item [3.]{Though OJ~287 is not in a rich cluster, a number of companion 
objects are in close proximity to OJ~287, some of which may be 
physical companions and three of which correspond with 
continuum radio emission seen near OJ~287.}

\item [4.]{{\it HST} imaging does not strengthen the case for an optical 
jet at P.A. 220 degrees.  The apparent alignment of objects in
ground-based images disappears when the field is observed with {\it HST}
resolution and there is no evidence of extended emission linking the distinct objects.
The apparent alignment of sources appears to be the result of a chance coincidence of
Galactic and extragalactic objects.}

\end {itemize}

Deeper high spatial resolution $R$ and $I$-band images of OJ~287 are
needed to search for optical counter parts to the radio emission seen
8-10$''$ west of OJ~287. Spectroscopy of objects A, E, F, and G is
needed to test the hypothesis that there is a merging group spread
over the 50 $h_{100}^{-1}$ kpc surrounding OJ~287 and potentially supplying
fuel to the AGN.

We thank Ron Kollgaard for several insightful comments.
We acknowledge support from NASA grant GO-5992.02-94A.
B.Y. acknowledges support from the Fermi National Accelerator Laboratory.

\vfil\eject
\centerline{\bf FIGURE CAPTIONS}

\bigskip
FIG. 1.--- WFPC2  was used to obtain $I$-band (F814W) images of the
BL~Lac object OJ~287 on 24 March 1996 UT.  This combined image, made
from a total of 1800 s of integration, shows the central
portion of the PC CCD image. The bright object is OJ~287. We have
followed and extended the object notation of B96. One
arcsecond is approximately 3 $h^{-1}$ kpc at z=0.306, the redshift of
OJ~287.

FIG. 2.--- PSFs:  both a) TinyTim model  and b) archival PSF from an
actual saturated star image (with the core fixed) in the same filter
and near the center of the WFPC-2 PC CCD.  These PSFs have excellent
S/N in their outer parts. The PSF in b) has been CR cleaned by hand.

FIG. 3.--- The {\it HST} image of OJ~287 described in Figure~1, but with a
scaled version of the PSF shown in Figure~2~a) subtracted from the image.  
The close companion object, E, and the diffuse emission to the West
and NW of the point source are more visible in this image.  The
candidate ``jet'' of Ben\'itez et al.~1996 at P.A. 220 degrees is
likely just the chance alignment of other objects. The arrows point
to object OJ-SW.

\begin{table}
\caption{Objects Near OJ~287 \label{tab1}}
\begin{tabular}{rrrrrr}
\tableline
\multicolumn{1}{c}  {ID} &
\multicolumn{1}{c} {$\Delta$RA($''$)} &
\multicolumn{1}{c} {$\Delta$DEC($''$)} &
\multicolumn{1}{c} {$I_{KC}$ mag} &   
\multicolumn{1}{c}  {FWHM\tablenotemark{a}} &
\multicolumn{1}{c}  {Ellip\tablenotemark{b}} \\
\tableline
OJ~287 &  0.00&   0.00&  $13.30\pm0.10$ & pt & $--$ \cr
OJ$-$Ext & $-$0.30&   0.30&  $18.30\pm  1.30$ & ext & yes  \cr
OJ$-$SW & $-$2.47&  $-$0.48&  $24.80\pm 1.00$ & $--$ & $--$  \cr
OJ$-$Gh\tablenotemark{c} & $-$2.63&  2.03&  $24.80\pm 1.00$ & $--$ & $--$  \cr
 A & $-$4.45&  $-$7.21&  $24.10\pm 0.50$ & pt & $--$  \cr
 B & $-$8.56&  $-$10.7&  $20.86\pm 0.05$ & pt & $--$  \cr
 C & $-$10.3&  $-$13.5&  $21.95\pm 0.20$ & pt & $--$ \cr
 E & $-$1.30&   3.30&  $20.62\pm 0.20$ & ext & 0.2  \cr
 F & $-$2.51&  10.09&  $:22.17\pm 1.00$ & ext & $--$ \cr
 G &  7.88&  $-$4.05&  $23.29\pm 0.50$ & ext & $--$ \cr
 H & $-$4.29&  $-$17.8&  $19.86\pm 0.20$ & ext & 0.6  \cr
 I &  7.28&  15.57&  $23.72\pm 0.50$ & ext & 0.8  \cr
 J & 10.29&  19.12&  $22.47\pm 0.30$ & ext & $--$  \cr
 K & 17.95&  $-$11.9&  $23.41\pm 0.50$ & ext & $--$ \cr
\end{tabular}
\tablenotetext{a}{ext indicates the object is significantly extended beyond a pt source}
\tablenotetext{b}{1-b/a axis ratio, for measurably elongated objects}
\tablenotetext{c}{Ghost image}
\end{table}

\end{document}